\documentstyle[10pt]{article}
\newenvironment{mytitle}{\begin{center} \LARGE}{\\ [.1in]\end{center}} 
\newenvironment{myauthor}{\begin{center} }{\\ [.1in]\end{center}} 
\newenvironment{myinstit}{\begin{center} \it}{\end{center}}

\begin{document}

\begin{mytitle}
Classical versus Quantum Structure of the Scattering Probability Matrix. 
Chaotic wave-guides.
\end{mytitle}
\vspace{.5cm}

\begin{myauthor}
G. A. Luna-Acosta$^1$, J. A. M\'endez-Berm\'udez$^1$, P. \v{S}eba$^{2,3}$, 
and K. N. Pichugin$^{2,4}$.
\end{myauthor}

\begin{myinstit}
$^1$Instituto de F\'{\i}sica, Universidad Aut\'onoma de Puebla, Apdo. Postal 
J-48, Puebla 72570, M\'exico.\\
$^2$Department of Physics, University Hradec Kralove, Czech Republic.\\
$^3$Institute of Physics, Czech Academy of Sciences, Cukrovarnicka 10, 
Prague, Czech Republic.\\
$^4$Kirensky Institute of Physics, 660036 Krasnoyarsk, Russia.
\end{myinstit}

\begin{abstract}
The purely classical counterpart of the Scattering Probability Matrix (SPM)  
$\mid S_{n,m}\mid^2$ of the quantum 
scattering matrix $S$ is defined for 2D quantum 
waveguides for an arbitrary number of propagating modes $M$. We compare the 
quantum and classical structures of $\mid S_{n,m}\mid^2$ for a waveguide 
with generic Hamiltonian chaos. It is shown that even for a moderate number 
of channels, knowledge of the classical structure of the SPM allows us 
to predict the global structure of the quantum one and, hence, understand 
important quantum transport properties of waveguides in terms of purely 
classical dynamics. It is also shown that the SPM, being an intensity 
measure, can give additional dynamical information to that obtained by the 
Poincar\`{e} maps.

\end{abstract}

\vspace{.5cm}

\section{Introduction}

The $S$-matrix is the most fundamental tool for analyzing quantum scattering 
phenomena in various fields of physics, for it provides us with the most 
complete scattering data \cite{TNN}. Moreover, it is often of 
interest to extend the analysis to the semiclassical regime. The first 
semiclassical formulations of the $S$-matrix appeared in the early 70s in the 
works of Miller \cite{Miller} and Marcus \cite{Marcus} in applications to 
atomic physics, and some years later extensions to their work were carried 
out by Heller \cite{Heller}. Their common approach uses the Feynman 
propagator in the WKB approximation, thus taking into account classical 
dynamics together with quantum mechanical interference, where the phases are 
given by the classical actions. In Ref. \cite{list} some important further 
developments on the semiclassical treatment of scattering systems are listed. 
Quantum and semiclassical calculations of the $S$-matrix have become 
essential for the understanding of transport phenomena in mesoscopic systems 
\cite{mesos}. In particular, in the ballistic regime, the conductance is 
well described by the Landauer-Buttiker formula \cite{Landauer} 
$G=\frac{2e}{\hbar^2}\sum_{n} \sum_{m}\mid t_{n,m} \mid^2$, where 
$t_{n,m}$ are the transmission elements of the $S$-matrix. Semiclassical 
expressions for the transmission amplitudes for collinear leads were obtained 
by Jalabert, Baranger, and Stone \cite{JBS}. See also Lin \cite{Lin}.\\

One of the aims of studying ballistic motion in  mesoscopic systems has been
to relate the experimentally observed behavior of transport quantities, in 
the classical and quantum regimes, to their underlying classical dynamics 
\cite{Several}. This is particularly interesting when the associated classical 
dynamics can be chaotic; then the purpose is to identify signatures of chaos 
in the transport \cite{Akguc}. In a very recent example of this kind of work Ketzmerick 
\cite{Ketz} showed that the fractal fluctuations of $G$ as a function of 
magnetic field in a chaotic cavity are related to the Poincar\`{e}-Birkhoff 
hierarchical structure of the phase space of the corresponding classical 
motion. Previously, Jalabert, Baranger, and collaborators performed 
detailed quantum and semiclassical calculations of conductance in mesoscopic 
systems that display chaos in the classical regime \cite{Bara}. An important 
conclusion from their work is that the behavior of the average conductance 
can already discern whether the underlying classical dynamics is regular or 
chaotic. Specially relevant for our purposes here is their finding 
that the dominant contributions to the average quantum conductance are 
classical \cite {JalaBara}. Thus it is natural to expect that useful 
information may be obtained by analyzing {\it purely} classical quantities, 
disregarding interference effects completely. Clearly, all the information 
contained in the phases of the quantum $S$-matrix, necessary to calculate 
e. g., the Wigner-Smith delay time, does not exist in the purely classical 
description. Nevertheless, as we shall show, important information can be 
extracted by studying the {\it Scattering Probability Matrix} (SPM); its 
elements are defined by the square modulo of the $S$-matrix elements 
$\mid S_{n,m}\mid^2$, which give the transition probability for an incoming 
mode $m$ to scatter into a mode $n$. The analysis of quantum and classical 
SP matrices is clearly relevant also for the study of the wave-ray 
correspondence of electromagnetic fields propagating in cavities \cite{doron} 
since, under certain conditions, the wave equations are the same as for the 
quantum ballistic transport \cite{Stockmann}.\\

In this paper we shall construct the purely classical counterpart of the 
quantum SPM valid for any 2D waveguide of arbitrary shape. Before doing so, 
in the next section we briefly review the definition of the $S$-matrix in its 
application to cavities connected to leads. In Sect. III we construct the 
classical SPM and compare its quantum and classical structures for a model 
of a mesoscopic ballistic 2D wave-guide that displays generic chaos in the 
classical limit. We shall show that the good global correspondence between 
classical and quantum SPM enables us to understand the classical dynamical 
origin of features of the quantum SPM and to clearly identify the differences 
produced by the wave nature of the quantum state. In Sect. IV, we make some 
concluding remarks.\\

\section{$S$-matrix for waveguides}

The $S$-matrix relates incoming waves to outgoing waves,

\begin{equation}
V^{out} = \hat{\bf S}\ V^{in},
\end{equation}

\noindent where $V^{in}$ and $V^{out}$ stand for vectors specifying, 
respectively, waves coming into and going out of the interaction region. 
For a system composed of a 2D wave-guide of arbitrary shape connected to two 
leads, say left (L) and right (R) leads, the solutions in the leads are 

\begin{equation}
\Psi^{L,R}(x,y)= \sum_{m=1}\left[a^{L,R}_m \exp(i k^{L,R}_m x) + 
b^{L,R}_m \exp(-i k^{L,R}_m x)\right]\phi^{L,R}_m(y),
\end{equation}

\noindent where

\begin{equation}
\phi^{L,R}_m(y)= \sqrt{\frac{2}{d_{L,R}}} \sin \left( \frac{m\pi y}{d_{L,R}}
\right) 
\end{equation}

\noindent is the component of the wave function along the $y$-axis 
perpendicular to the direction of propagation ($x$-axis); $d_L$ stands for 
the constant width of the left lead which may be different from the width 
$d_R$ of the right lead. For simplicity's sake we shall use $d_L=d_R$ for 
the rest of the paper. 
The sum is over all the propagating modes supported 
by the leads at a given Fermi Energy $E$.\\

With this notation the $S$-matrix and the incoming and outgoing waves can be 
written in the form 

\begin{eqnarray}
\hat{\bf S} = \left(
\begin{array}{ll}
t & r' \\
r & t 
\end{array}
\right ),\ \
V^{in}= \left(
\begin{array}{l}
a^L \\
b^R \\
\end{array}
\right),\ \
V^{out}=
\left(
\begin{array}{l}
a^R\\
b^L
\end{array}
\right).\nonumber
\end{eqnarray}

The symbols $t$ ,$t'$, $r$, and $r'$ in the $S$-Matrix are $M \times M$ 
matrices, where $M$ is the highest mode (given by the largest $m$ beyond 
which the longitudinal wave vector 
$k^{L,R}_m =\sqrt{\frac{2E}{\hbar^2}-\frac{m^2 \pi^2}{d_{L,R}^2}}$ becomes 
complex). The symbols $a^{L,R}$ and $b^{L,R}$ stand for the vectors $a^{L,R}_m$ and 
$b^{L,R}_m$, $m=1,2...M$. The {\it squared} modulo element $\mid 
t_{n,m}\mid^2$ ($\mid t'_{n,m}\mid^2$) gives the probability amplitude for a 
left (right)-incoming mode $m$ to  be transmitted to the right (left) lead 
into the mode $n$. Similarly, $\mid r_{n,m}\mid^2$ ($\mid r'_{n,m}\mid^2$) 
is the probability for a left (right)-incoming mode $m$ to be reflected to 
the left (right) lead into mode $n$. \\

The quantum SPM is simply defined as $\mid S_{n,m}\mid^2$; it gives the 
transition probability for the incoming mode $m$ to transmit or reflect into 
an outgoing mode $n$.\\

\section{Classical Scattering Probability Matrix (SPM)}.

Since the energy of the system is given by its expression in the leads

\begin{equation}
E=\frac{\hbar^2}{2m_e} \left( k_m^2 +\frac{m^2 \pi^2}{d^2} \right),
\end{equation}

\noindent classically we can associate an angle $\theta_m$ between the 
longitudinal component of the momentum $k_m$ and the total momentum 
$\sqrt{2m_eE}/\hbar$. That is,

\begin{eqnarray}
\theta_m = \sin^{-1} \left[ \frac{m \pi \hbar}{d\sqrt{2m_eE}} \right].
\end{eqnarray}

For a finite number $M$ of modes there corresponds a range of angles 
$\Delta \theta_m\equiv\theta_m-\theta_{m-1}$ for each mode $m$. The classical 
limit is $M=\infty$.\\

Consider a classical particle entering, say, from the left side and making an 
angle $\theta_i$, within a range corresponding to a given mode $m$. The 
particle (ray) will generally collide with the walls of the wave-guide a few 
times before exiting to the left or to the right, making a certain angle 
$\theta_f$, to which we can associate a mode $n$ if 
$\theta_f \in \Delta \theta_n$. Initial conditions are specified not just by 
the angle but also by the initial position $(x,y)$ along the left lead. In 
order to account for all possible types of trajectories, we take a large 
number of 
initial positions for each incoming angle $\theta_i$. By recording the number 
of particles scattered into the various ranges of $\theta$ associated with 
different outgoing modes ${n}$, we obtain a distribution of outgoing modes 
for each incoming mode $m$. This distribution gives the {\it classical counterpart} 
of the matrix elements $\mid t_{n,m} \mid^2$ and $\mid r_{n,m} \mid^2$ of the 
quantum SPM. Similarly, to obtain the classical counterparts of 
$\mid t'_{n,m} \mid^2$ and $\mid r'_{n,m} \mid^2$ we repeat the above process 
but for particles entering from the right lead. This defines the procedure to 
construct the {\it classical counterpart of the SPM}.\\

{\bf The wave-guide}\\

We now specify a wave-guide model on which to explore the quantum and 
classical structures of the SPM. We chose the geometry of the wave-guide to 
be that of a ``rippled" billiard, shown in Fig. 1, which is connected to two 
collinear leads of the same width. We have selected this particular shape for 
the wave-guide because it is known to display all features of an important 
class of dynamical systems, namely chaotic billiards that undergo the generic 
(Hamiltonian) transition to chaos \cite{Licht,gala1}.\\
 
Moreover, the finite version, depicted in Fig. 1, which serves as a model of 
a quantum or electromagnetic waveguide, has been used to study certain 
transport manifestations of chaos in the classical \cite{gala1} as well as 
quantum \cite{ketz2} regimes. On the other hand, the infinitely long ({\it 
i.e.}, periodic) version of the rippled billiard, introduced first in 
connection to beam acceleration problems \cite{Month}, has been useful also 
for the understanding of typical features of crystals (e. g., energy band 
structure, LDOS, etc.) and their quantum-classical correspondence 
\cite{gala2,gala3,gala4}.\\
   
Although, as a scattering system the finite version of the rippled channel is 
the relevant one, it is convenient first to review briefly the motion in the infinitely long 
rippled billiard, $L\rightarrow \infty$. As usual, to get the dynamical 
panorama, we look at a Poincar\`{e} map of the system. As the Poincar\`{e} 
surface of section we choose, for reasons of symmetry, the bottom boundary 
$y=0$; the Poincar\`{e} map is given by the pair of Birkhoff variables 
$(x_j,\theta_j)$, labeling the longitudinal components of the position and 
angle of the particle right after its $j^{th}$ collision with the bottom 
wall. Since the channel is periodic the Poincar\`{e} map is on the cylinder 
({\it i.e.}, $x$ is mod. $1$). Depending on the 
geometrical parameters (average width $\gamma$ and amplitude $\nu$ of the 
ripple) the dynamics is either regular, mixed, or fully chaotic. Figs. 2a and 
2b show, respectively, typical Poincar\`{e} sections for a wide ($\gamma = 0.5$, 
$\nu = 0.12$) and a narrow ($\gamma = 0.25$, $\nu = 0.025$) channel. 
In general, for small amplitudes of the ripple ($\nu<<1$) wide channels 
($\gamma \stackrel{>}{\sim} 1/2$) give rise to global chaos, whereas narrow 
channels yield mixed dynamics, as exemplified by Figs. 2a and 2b. For future 
reference we shall denote the system displaying globally chaotic dynamics 
($\gamma = 0.5$, $\nu = 0.12$) as the {\it G system} and the mixed one 
($\gamma = 0.25$, $\nu = 0.025$) as the {\it M system}.\\

Since the Poincar\`{e} plots of the {\it periodic} rippled billiard show 
topological chaos ({\it i.e.}, a heteroclinic tangle), it is not surprising 
that a {\it finite} rippled billiard connected to leads shows chaotic 
scattering, as evidenced by the fractal nature of its scattering functions, 
such as the dwelling time, reported in \cite{gala1}. In fact, as is well 
known \cite{EckSmi}, topological chaos is responsible for the fractality of 
the scattering functions.\\

\section{Results}

In the following we shall compare the quantum and classical SP matrices for 
both systems, {\it G} and {\it M}, and for various lengths of a rippled 
wave-guide. In all cases we will consider energies that allow for 33 
propagating modes.\\

{\bf The G-system}\\

Fig. 3a and Fig. 3b show, respectively, the quantum and classical SPM for the 
rippled wave-guide whose length equals {\it one} period of the ripple 
($L = 2 \pi$). The resemblance between the classical and quantum SP matrices is 
remarkable. Let us consider first the reflection part of the SP matrices, say 
the left bottom block $\mid r_{n,m}\mid ^2$. Notice in the classical SPM the 
high intensity in the neighborhood of the $(n,m)=(11,11)$ element and along a 
hyperbola like curve centered on it. The quantum SPM also shows the same 
pattern. The same is true for the cone-like shape starting at the 
$(n,m)=(11,11)$ element. More impressive is the similitude of triangular shapes near 
the top right corner. This global correspondence enables us to predict, based 
solely on the classical pictures, important quantum transport features. For 
example the classical SPM predicts that there will be negligible reflection 
for modes $m\leq 4$. This is confirmed in Fig. 4, which 
shows in detail how the incoming modes 2 and 3 do not reflect, whereas the 
incoming mode number 5 does reflect partially into the outgoing mode number 
$24$, just as predicted classically [note the high intensity element 
$(n,m)=(24,5)$]. These figures also show that the modes 2, 3 and 5 transmit 
predominantly onto the same modes as the incoming ones; this would be just 
like the classical probabilities except that the quantum one shows, in 
addition, small transmission to some modes off the diagonal. Detailed 
analysis of the data shows that the classical SPM also gives transmission off 
the diagonal but it is not evident because their intensity is weak and almost 
uniform over all modes. 
This difference is due to quantum interference effects, which are 
responsible too for the larger width of the diagonal elements of the 
transmission parts. As another example, the classical SPM predicts that mode 
$11$, incoming from the left, will reflect and transmit predominantly onto 
the same channel number, which is confirmed by Fig. 4b.\\

It is instructive to identify the type of trajectories that form the most 
salient features of the classical SPM since these are also evident in the 
quantum SPM. As an illustration, the triangular shape (see Fig. 5) that appears near the 
top right corner of the $\mid r_{n,m} \mid^2$ block results from incoming 
trajectories colliding only {\it once} with the rippled boundary in the 
neighborhood of $x=1/2$; the hyperbola-like curve and also the cone-like 
shape are formed by trajectories colliding {\it twice} with the rippled 
boundary.\\

An important aspect of the quantum-classical correspondence which was not 
expected is the particle-like behavior that results from the interaction with 
the rippled wave-guide of certain plane waves. As an example, Fig. 6 illustrates 
this behavior for the incoming waves with mode numbers 24 and 29. Note that a high 
intensity pattern is formed on the left side of the rippled wave-guide, 
resembling a ray trace. The angle this pattern makes with the horizontal, 
labeled $\alpha$ in the figure, corresponds precisely with the angle of 
reflection predicted by the classical SPM. In general we see that when there 
is a high intensity element in the classical SPM one can expect the wave 
function to form a ray pattern along the classical trajectory just outside 
the cavity. This may be regarded as a ``short-lived scar".\\ 

In contrast, when the classical SPM shows {\it homogeneous areas of 
low intensity} probabilities, the quantum SPM is expected to show a mottled 
pattern of medium intensity probabilities. A homogeneous area of low 
intensity classical probabilities results when 
incoming particles within a range $\Delta \theta_m$ scatter uniformly 
throughout a much wider range of angles. This effect, the defocusing caused 
by the rippled boundary, is responsible for the strong sensitive dependence 
to initial conditions, the main ingredient of chaos. Clearly, the larger the 
number of periods forming the rippled wave-guide, the stronger this effect 
should be. Figs. 7a and 7b, showing the quantum and classical SP matrices for 
the same geometry as just above ($\gamma = 0.5$, $\nu = 0.12$) , but six 
times longer ($L = 12 \pi$), confirms this expectation for the {\it 
transmission} parts. Comparison of the classical SP matrices (Figs. 3b and 
7b) shows that, with the exception of a 
high spot near the transmission element $(n,m)=(28,28)$ and a short diagonal 
contribution $(n,m)<(5,5)$, all the distinguishing features of the 
transmission parts obtained for a one-period-long wave-guide (Fig. 3b) are 
washed out in the case of the six-period long wave-guide. The remaining high 
intensity diagonal elements for low modes are due to direct transmission, 
{\it i.e.}, to trajectories that transmit without colliding with the upper wall. 
To get an estimate for the number of incoming modes that transmit 
predominantly onto the same mode, assume flat boundaries (since the 
amplitude of the ripple is small compared to the width $\gamma$) and consider 
a bundle of particles injected at $(x,y)=(0.0,0.5)$. The particles can 
transmit directly (no collisions with upper or lower boundary) if their 
initial angle $\theta_i$ is in the interval $(-\theta_c,\theta_c)$, where 
$\theta_c\equiv tan^{-1}(2L/d)$, and $L$ is the length of the channel. 
For the one-period-long wave-guide and six-period-long wave-guide, these 
angles are, respectively, $0.46$ and $0.083$ radians. Their ratio is $0.18$, 
which agrees with the ratio between the lengths of the high intensity 
diagonals in the transmission of Figs 3b and 7b.\\

We remark that the homogeneous spread of intensities in the transmission 
part of the classical SPM 
for $(n,m)>(5,5)$, is consistent with the ``equal {\it apriori} distribution" 
of the $S$-matrix required for the validity of the random $S$-matrix theory 
approach to chaotic cavities \cite{BaMe94}. But note that in the reflection 
parts of the classical SPM the inhomogeneity is especially strong. In 
fact, while the definite transmission structures of the one-period wave-guide 
have been somewhat washed out in the six-period wave-guide, {\it classically}, the 
reflection blocks remain practically the same as for the one-period-long 
wave-guide. This is because the reflection structures are mainly formed by 
particles reflecting within the first period of the ripple, see Fig. 5. On 
the other hand, backscattering after the first period of the rippled 
wave-guide shows up classically as an almost homogeneous spread of intensity 
throughout the reflection blocks (including the area below the hyperbola-like 
curve for which there was no reflection for the one-period wave-guide). Thus, 
while classically the definite pattern produced by the first period of the 
wave-guide persists, quantum interference of the backscattering from the 
whole wave-guide starts to destroy the pattern observed in the classical 
SPM.\\
 
Another interesting feature comes from the analysis of the relatively bright 
spot observed in the transmission part of the classical SPM near the site 
$(n,m)=(28,28)$. A zoom of this spot is shown in Fig. 8a and a typical 
trajectory belonging to this pattern is shown in Fig. 8c. The distinctive 
feature of this type of trajectories is that they collide twice with the 
rippled boundary for each bounce with the flat boundary. These are periodic 
or quasiperiodic orbits advancing always to the right and form the stability 
island surrounding the stable period one fixed point shown in Fig. 8b. It is 
important to remark that this miniscule Poincar\`{e}-Birkhoff structure (note 
the scale of the axis) is not visible in the whole Poincar\`{e} map of Fig. 2a, 
even though its effect is clearly manifested in the classical SPM. Hence, we 
see that the SPM construction can give complementary information to that 
obtained by the Poincar\`{e} maps because it is an intensity measure. On the 
other hand, such spot is not present in the quantum SPM because the size of 
the Poincar\`{e}-Birkhoff structure is too small to be resolved 
quantum-mechanically (there are other spots visible but they do not 
correspond to the classical one; they originate from constructive 
interference).\\

{\bf The M-system}\\

Now we examine briefly the classical and quantum SP matrices for the $M$ 
system, Figs. 9a and 9b show these matrices for a one-period-long wave-guide. 
Again, a quick comparison of these shows that the global features of the 
quantum SPM can be predicted by the classical SPM. We see that regions of 
high intensity areas in the quantum SPM correspond roughly to the areas 
of the classical SPM, albeit fluctuations within them. However, there are 
some important differences which we shall discuss now. Note that both 
classical and quantum SP matrices show that reflection occurs only for high 
modes but the classical reflection occurs only for modes higher or equal to 
$m=31$ while quantum reflection (although weak) exists even for modes as low 
as $m=10$. The mechanism responsible for the reflection of classical 
particles can be understood by examining the Poincar\`{e} map of the 
infinitely long channel, Fig. 2b, which 
shows a large resonance island centered at $x=\frac{1}{4}$. This resonance 
is produced by trajectories executing librational motion, bouncing between 
the two walls in the neighborhood of the widest part of the channel, 
$x=\frac{1}{4}$. It is clear then that particles entering the rippled 
wave-guide from the left at $x=0$ can reflect (after one or several bounces) 
within the {\it first period} of the channel if their trajectories fall 
within the resonance island. Trajectories falling on the chaotic sea outside 
the resonance island (hence low transversal mode numbers) can also reflect 
via the chaotic separatrix but not within the first period of the ripple. 
The longitudinal momentum of 
these librational orbits is relatively small (see Fig. 2b), hence their 
transverse momentum is large. Detailed analysis using equation (5) and data 
from Fig. 2b shows that indeed the lowest mode that can reflect is $m=31$, 
in agreement with Fig. 9b. In contrast there is a strong quantum reflection 
for modes as low as $m=25$. Heisenberg's Uncertainty Principle is responsible 
for this difference, namely, the quantum 
state cannot resolve the fine classical boundaries defining the resonance 
island and consequently even smaller values of $m$ can ``tunnel" into the 
resonance island to cause partial reflection.\\

\section{Concluding Remarks}

We have studied quantum scattering properties of typical wave-guides with 
mixed and global chaos by examining the quantum 
{\it Scattering Probability Matrix} (SPM) and its classical counterpart. We 
emphasize that the definition of the classical SPM does not include any 
semiclassical aspects. We showed that the structure of the classical SPM 
allows us to predict the global structure of the quantum SPM. Since features 
of the classical SPM can be understood by analysis of the trajectories, it 
was possible to understand the classical dynamical origin of important 
features of the quantum SPM. Consequently, the analysis of the classical SPM 
of a given electron wave-guide system is useful for the understanding of 
its quantum transport properties, e. g., conductance. Plots of the classical 
SPM can be examined quickly to determine the influence of the cavity on the 
various modes. For a given energy, some modes may show ballistic behavior 
while others may display diffusive transmission, as observed recently in 
Ref. \cite{Sanchez}.\\

Our analysis of the quantum-classical correspondence of the SPM led us to 
discover the existence of ``short-lived scars". Specifically, we have seen 
that the wavefunction forms a ray pattern along the outgoing classical 
trajectory, for modes corresponding to high-intensity elements of the 
classical SPM. They are short-lived because after few bounces the ray pattern 
is destroyed by quantum interference.\\

Finally, we wish to mention that certain small but relatively high intensity 
areas in the classical SPM lead us to discover the existence of extremely 
small Poincar\`{e}-Birkhoff structures of the otherwise globally chaotic 
billiard. Hence, the SPM gives complementary information to that obtained 
solely by topological tools (e. g., Poincar\`{e} maps) since it is an 
intensity measure.\\   

{\bf Acknowledgements}: We wish to acknowledge financial support from 
CONACYT (Mexico) grant No. 26163-E.

\newpage

\begin{center}
{\Large \bf FIGURE CAPTIONS}
\end{center}
\vspace{.5cm}

{\bf Fig. 1} Geometry of the wave-guide.\\

{\bf Fig. 2} Poincar\`{e} surface of section at $\gamma=0$ for (a) 
$(\gamma,\nu)=(0.5,0.12)$ and (b) $(\gamma,\nu)=(0.25,0.025)$.\\

{\bf Fig. 3} (a) Quantum and (b) Classical SPM, $|S_{n,m}|^2$, for the 
one-period wave-guide with $(\gamma,\nu)=(0.5,0.12)$.\\

{\bf Fig. 4} Reflection and transmission probabilities from the classical SPM 
for (a) the incoming modes 2, 3, 5, and (b) 11. The one-period wave-guide 
with $(\gamma,\nu)=(0.5,0.12)$ is considered.\\

{\bf Fig. 5} Typical types of trayectories that contribute to the zones 
marked in the reflection part $|r_{n,m}|^2$ of the classical SPM of Fig. 3.\\

{\bf Fig. 6} Wave function of the incoming mode (from the left) number (a) 
24, and (b) 29. (a) $\alpha \sim 60^\circ$, and (b) $\alpha \sim 47^\circ$ 
are the reflection angles predicted by the classical SPM of Fig. 3.\\

{\bf Fig. 7} (a) Quantum and (b) Classical SPM, $|S_{n,m}|^2$, for the 
six-period wave-guide with $(\gamma,\nu)=(0.5,0.12)$.\\

{\bf Fig. 8} (a) Zoom on the transmission part of the classical SPM arround 
the site $(n,m)=(28,28)$. (b) Phase space generated by the trajectories that 
produce the structure of (a). (c) Typical orbit belonging to the pattern in
(c).\\

{\bf Fig. 9} (a) Quantum and (b) Classical SPM, $|S_{n,m}|^2$, for the 
one-period wave-guide with $(\gamma,\nu)=(0.25,0.025)$.\\

\end{document}